\documentclass[aps,prx,reprint,twocolumn,superscriptaddress,nofootinbib,floatfix]{revtex4-2}
 
\usepackage{float}
\usepackage{verbatim}

\usepackage[english]{babel}
\usepackage{dsfont}
\usepackage[normalem]{ulem}
\usepackage{amsmath}
\usepackage{amssymb}
\usepackage{mathrsfs}
\usepackage{amsthm}
\usepackage{mathtools}
\usepackage{wasysym}
\usepackage{breqn}
\usepackage{tabularx}
\usepackage{bbm}
\usepackage{nicefrac}
\usepackage{soul}
\usepackage{stackrel}
\usepackage{braket}
\usepackage{enumitem}
\usepackage{graphicx}
\usepackage{tikz}
\usepackage{mdframed}
\usepackage{booktabs}
\usepackage{siunitx}
\usepackage{fancyhdr}
\usepackage{overpic}
\usepackage{calc}
\usepackage{accents}

\usepackage{multirow}
\usepackage{array}
\newcolumntype{P}[1]{>{\centering\arraybackslash}p{#1}}

\usepackage[hidelinks]{hyperref}
\usepackage{framed}

\addto\captionsenglish{}

\usepackage{cases}

\addto\captionsenglish{}
\addto\captionsenglish{}

\usepackage{algorithm}



\usepackage{natbib}
\makeatletter
\let\cat@comma@active\@empty
\makeatother

\begin{document}
\title{On-demand storing time-bin qubit states with optical quantum memory}

\author{Ming-Shuo Sun}
\author{Chun-Hui Zhang}
\author{Yi-Zhen Luo}
\affiliation{Institute of quantum information and technology, Nanjing University of Posts and Telecommunications, Nanjing 210003, China.}
\affiliation{ Telecommunication and Networks National Engineering Research Center, NUPT, Nanjing 210003, China.}
\author{Shuang Wang}

\affiliation{ Key Laboratory of Quantum Information, University of Science and Technology of China, CAS, Hefei 230026, China}
\author{Yun Liu}
\affiliation{ Anhui Asky Quantum Technology Co., Ltd., Wuhu 241002, China.}
\author{Jian Li}
\author{Qin Wang}\email{qinw@njupt.edu.cn}
\affiliation{Institute of quantum information and technology, Nanjing University of Posts and Telecommunications, Nanjing 210003, China.}
\affiliation{ Telecommunication and Networks National Engineering Research Center, NUPT, Nanjing 210003, China.}

\begin{abstract}

\noindent	Quantum memory, serving as a crucial device for storing and releasing quantum states, holds significant importance in long-distance quantum communications. Up to date, quantum memories have been realized in many different systems. However, most of them have complex structures and high costs. Besides, it is not easy to simultaneously achieve both high storage efficiency and fidelity. In this paper, we experimentally demonstrate a low-cost optical quantum memory with high efficiency and high fidelity, by utilizing a butterfly shaped cavity consisting of one polarization beam splitter, two reflecting mirrors and one pockels cell crystal. In order to quantify the quality of the quantum memory, we carry out tomography measurements on the time-bin qubits encoded with weak coherent states after storage for N rounds. The storage efficiency per round can reach up to 95.0\%, and the overall state fidelity can exceed 99.1\%. It thus seems very promising for practical implementation in quantum communication and networks.
\end{abstract}

\maketitle

\section{Introduction}
Achieving long-distance quantum communication remains a challenging task, as one of the most significant obstacles is the exponential optical attenuation over long distances. In quantum key distribution (QKD) \cite{BB84}, especially in measurement-device-independent QKD \cite{mdi,mdi2}, two-photon coincidence is essential for long-distance communication. To improve it, one way is to achieve the asynchronous matching of click results \cite{amdi1,amdi2,amdi3}, and the other way is to design an on-demand quantum memory that stores one photon until the other photon arrives \cite{QM2,QM}.
Nevertheless, the pursuit of an ideal quantum memory is full of challenges, primarily due to limitations in memory efficiency, the fidelity of retrieved states, and the lifetime of quantum states.  
\par
In an effort to overcome these challenges, a large number of schemes on quantum memory have been put forward. Considering diverse ensemble systems, numerous memories are founded on cold-atom ensembles \cite{cold1,cold2,obi}; ion-trap systems \cite{trap1,trap2}, which have achieved an ion-photon entanglement fidelity of 0.81 after 10 seconds \cite{trap1}; and solid-state systems \cite{solid-1,solid-2,solid-3,chip2}. Furthermore, these schemes also concentrate on specific encoding states. For instance, there are storage efficiencies of 85\% for 2-dimensional heralded single photon source polarization qubits \cite{pol},  single-photon level weak coherent states \cite{wcs,wcs2},  and single-photon level orbital angular momentum states \cite{obi}. To increase the information capacity, one scheme focusing on high-dimensional states has been constructed, achieving up to 25-dimensional perfect optical vortex modes \cite{HD25}. Another scheme is multiplexed quantum memory, typically implemented using multiple degrees of freedom of photons \cite{MUL1,MUL2}. Moreover, to enhance the stability of quantum storage and its scalability in quantum networks, significant progress has also been made in miniaturization and chip integration \cite{chip2,chip3,chip4}. In fact, chip-based memory can achieve 99\% fidelity of the retrieved time-bin states \cite{chip2}. However, the above memory schemes generally have high requirements on equipment and cannot simultaneously meet the demands of high efficiency, high fidelity, and flexible storage time.\par
Here, we present an on-demand all-optical quantum memory scheme that employs only one polarization beam splitter (PBS), two reflecting mirrors, and one pockels cell crystal (PC) \cite{Pock}. The high-reflective mirrors play a crucial role in ensuring the efficiency of memory and the lifetime of quantum states. The notable characteristic of a pockels cell is that when a high-level voltage is applied, the polarization of the transmitted light rotates by $90^{\circ}$. Leveraging this property, we can actively control the storage and read-out of the quantum state. To evaluate the performance of our cavity, we measure the memory efficiency and the lifetime of the stored state. Additionally, we conduct tomography measurement on the output state to determine its fidelity. Through the utilization of these technologies, we have achieved a quantum memory with an average single round efficiency of 95.0\%, a lifetime of 40 rounds, and a fidelity of over 99.1\% for all rounds of memory.

\section{An on-demand all-optical quantum memory}
\label{sec:b}
The scheme on the storage of time-bin qubit states, based on all-optical components, is illustrated in Fig. 1. A pulsed laser emits pulsed light at a frequency of 100 kHz with a temporal width of 50 ps. An attenuator is used to adjust the intensity. The FMI (Faraday-Michelson Interferometer), composed of two faraday mirrors (FMs) and one phase modulator (PM), is employed to generate  time-bins with a 1.5 ns time interval. The PM and the following intensity modulator (IM) are employed to encode the pulsed light into one of the six states, including Z basis: $|e\rangle$ and $|l\rangle$, X basis: $  |+\rangle=\dfrac{|e\rangle+|l\rangle}{\sqrt{2}}$ and $  |-\rangle=\dfrac{|e\rangle-|l\rangle}{\sqrt{2}}$, and Y basis:$ |L\rangle=\dfrac{|e\rangle+i|l\rangle}{\sqrt{2}}$ and $  |R\rangle=\dfrac{|e\rangle-i|l\rangle}{\sqrt{2}}$. Subsequently, the encoded states enter the optical cavity and are stored therein by applying a voltage to the pockels cell. 
The entire cavity consists of the following components: two high-reflective mirrors with a reflecting efficiency of 99.5\%; one pockels cell having a transmission efficiency of 98.3\%, a central wavelength of 1550 nm, and a bandwidth larger than 100 nm; one HWP; and one PBS. The extinction ratios of the PBS, the HWP, and the pockels cell all exceed 30 dB. The total length of one round of the cavity is approximately 4 meters.\par
We can apply another voltage to the pockels cell to read out the time-bin state. Then, the released states enter the second FMI for decoding purposes. Eventually, a single-photon detector (SPD) is utilized to detect the output photons, and a time-to-digital converter (TDC) is used to record the time information.\par

\begin{figure*}[t]
	\centering
	\includegraphics[scale=0.18]{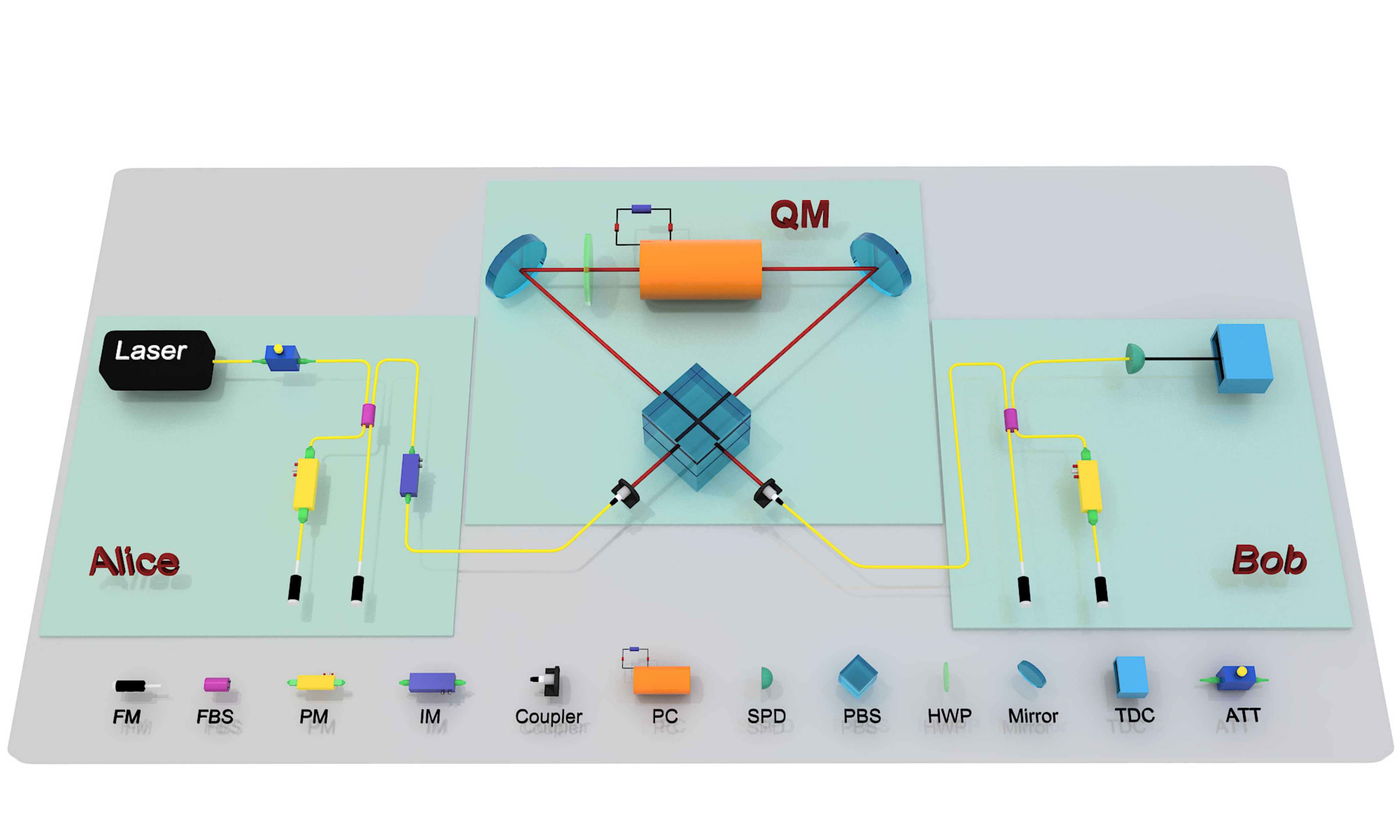}
	\caption{The schematic of the quantum memory setup. FM: faraday mirror; FBS: fiber beam-splitter; PM: phase modulator; IM: intensity modulator; PC: pockels cell;  SPD: single-photon detector; PBS: polarization beam splitter; HWP: half wave plate; Mirror: high reflectivity mirrors; TDC: time-to-digital converter; ATT: attenuator}
	\label{Fig1}
\end{figure*}

To realize the on-demand readout of quantum states, an appropriate setup is necessary. We set the polarization of the light emitted from the laser to be horizontal (H) and configure the HWP at an angle of $45^{\circ}$. At the PC, the gaussian beam waist radius of light is adjusted to meet the self-reproducing condition of the cavity. The high voltage signal of the PC, the trigger signal of the pulsed laser, and the single-photon detector are synchronized with the same clock. The detailed control sequence is depicted in Fig. 2. If quantum state storage is not needed, without any operation, the HWP will rotate the polarization twice. Due to the transmission property of the PBS, the light will be released.\par 
\begin{figure}[htbp]
	\centering
	\includegraphics[scale=0.6]{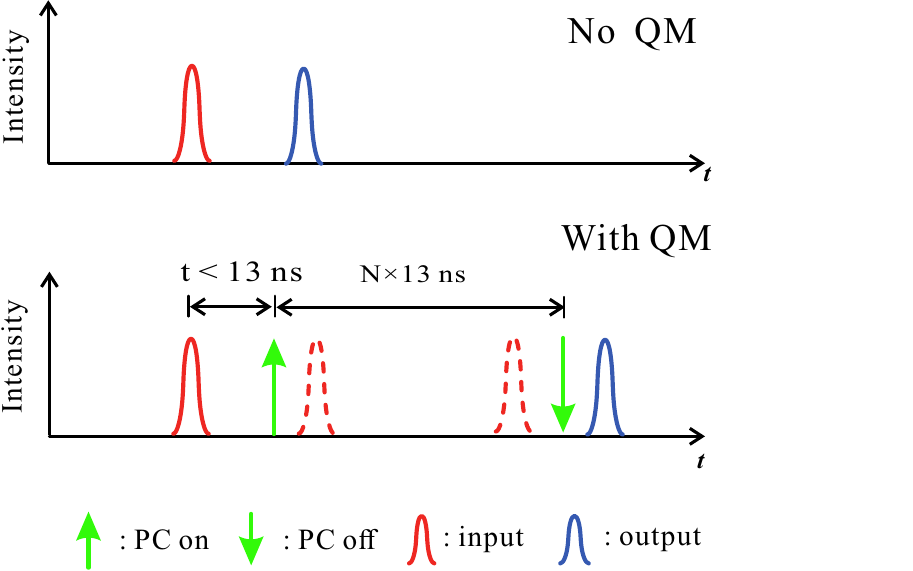}
	\caption{The control sequence of quantum storage. Without memory: When the pulse reaches the PC, without any operation, due to the property of the PBS, the pulse directly leaves the cavity. With memory: When the pulse reaches the PC, there is no high-voltage signal loaded on the PC. Within one period of the cavity i.e. 13 ns, a high-voltage signal is loaded on the PC. Eventually, the storage time in the cavity is equivalent to the pulse width of the high-voltage signal. Here, the solid pulse denotes the horizontal polarization and the dashed pulse denotes the vertical polarization.}
	\label{Fig1}
\end{figure}

For quantum state storage, the control sequence proceeds as follows. The input pulsed light arrives at the PC, which is initially loaded with a low-level signal. The polarization of the pulsed light is then changed from H to V by the following HWP.  Subsequently, within 13 ns, a high-level signal is loaded into the PC. Under this condition, both the PC and the HWP induce the polarization to change in the sequence of V - H - V. Owing to the reflectivity of the PBS, the pulsed light is stored within the cavity. When the high-level signal terminates, the polarization is altered from V to H by the HWP, and the pulsed light is released. The storage time is correlated with the width of the high-level signal, which is typically an integer multiple of the cavity period, $i.e.$, $N\times13$ ns.\par
Specifically, during the process of measuring the efficiency of quantum memory, we carefully collect the counts of different rounds ranging from 3 to 40, as depicted in Fig. 3. Here, the background noise mainly come from the dark counts 15Hz and the leaked light from HWP, PBS and PC. The signal-to-noise ratio (SNR) are 24.5dB at the 3$rd$ round and 21.9dB at the 30$th$ round. Subsequently, we fit these collected counts of different rounds to an exponential function of the form $\beta e^{-\alpha n}$.  It is worth noting that through comprehensive evaluation and calculation of the data, we have determined that $\alpha = 0.051$ and that the average efficiency is approximately 95.0\%. \par
\begin{figure}[htbp]
	\raggedright
	\includegraphics[scale=0.3]{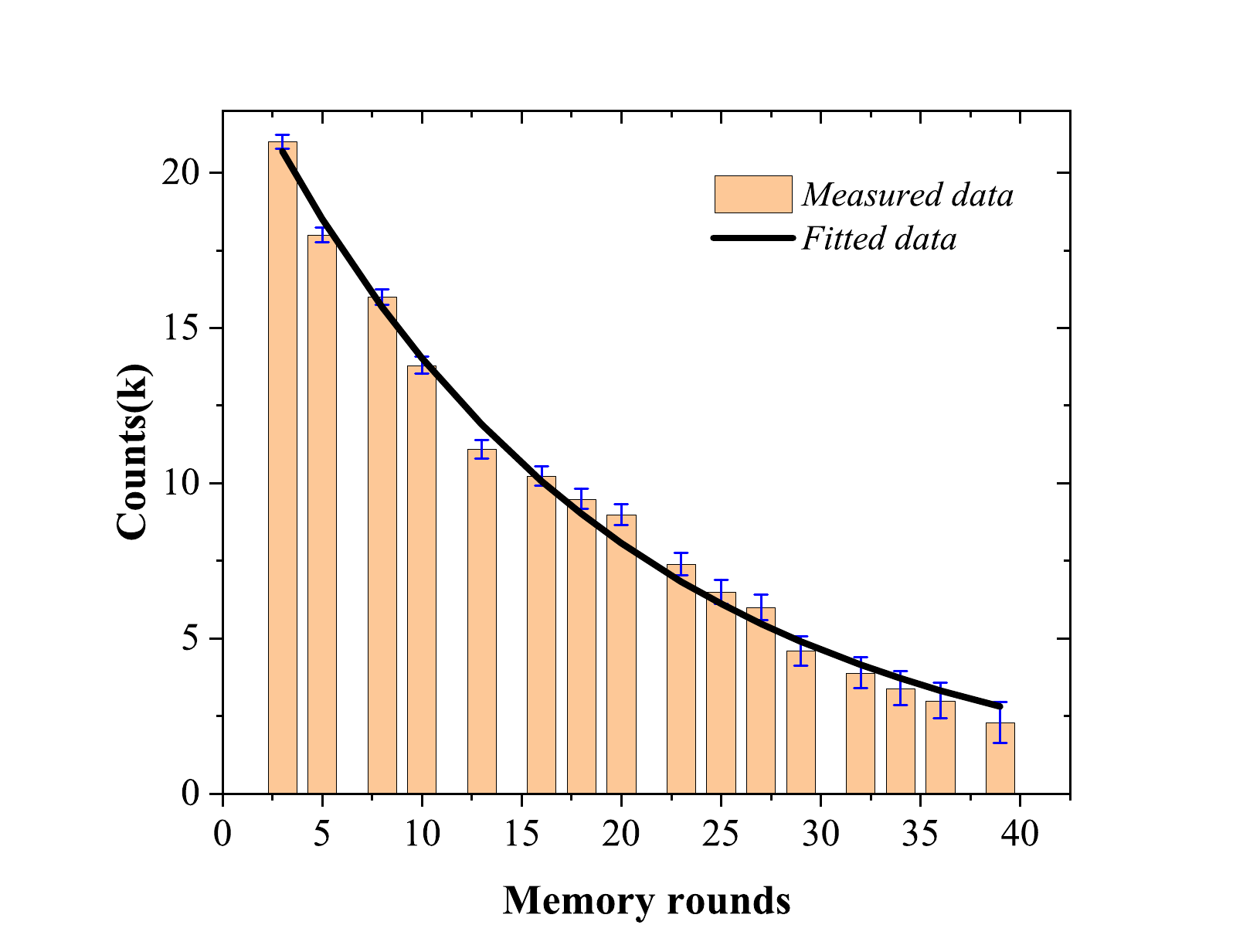}
	\caption{The counts of actual measurement and fitted exponential function. The column chart with error bar denotes the actual counts of different storage rounds. The solid line denotes the fitted counts of different storage rounds from our measured datas. The photon counts of one single measurement are collected for 1 s with 1 ns detection window.}
	\label{Fig1}
\end{figure}

\section{Tomography experiment of timebin-phase states quantum memory}
Tomography \cite{tomo} is a singularly helpful tool to estimate the density matrix $\rho$ of a quantum state. Let $\rho$ represents the density matrix of an arbitrary single-qubit state, which can be expressed as, $\rho=\dfrac{1}{2}(I+\overrightarrow{r}\overrightarrow{\sigma})$, here $\overrightarrow{\sigma}=(\sigma_{x}, \sigma_{y}, \sigma_{z})$, and $\sigma_{i}$ represents Pauli operators along each mutually perpendicular direction. Here, $\overrightarrow{r}=(r_{x}, r_{y}, r_{z})$ denotes the local Bloch vector. We measure different $\sigma$ with $n$ $(n>>1)$ times and corresponding outcomes will be denoted as $\{x_{i}=\pm1\}_{i=1,2,...,n}$. As  $r_{x}=Trace(\sigma_{x}\rho)=lim_{n\rightarrow \infty} \dfrac{\sum_{i=1}^{n}x_{i}}{n}$. Similarly to get other pauli operators, then the Bloch vectors $\overrightarrow{r}$ could be estimated.\par

We start the single-photon qubit storage measurement to confirm the relative phase is preserved during the storage process. Subsequently, from the measured matrices, the fidelity of the output states compared to the input ones could be estimated. It is defined as $F=(Trace[\sqrt{\sqrt{\rho_{in}}\rho_{out}\sqrt{\rho_{in}}}])^{2}$, where $\rho_{in}(\rho_{out})$ is the density matrix of the input (output) state. 
In Fig. 4, we calculate the average fidelity of memory round from 0 to 40. The fidelity of all storage rounds is over $99.1\%$ which is a comparatively good result. The error bars of high storage rounds are higher than those of the low storage rounds which may result from the reduction in data volume caused by attenuation. And, all the six reconstructed density matrices of the 30 $th$ memory round are shown in Fig. 5.\par

\begin{figure}[htbp]
	\includegraphics[scale=0.3]{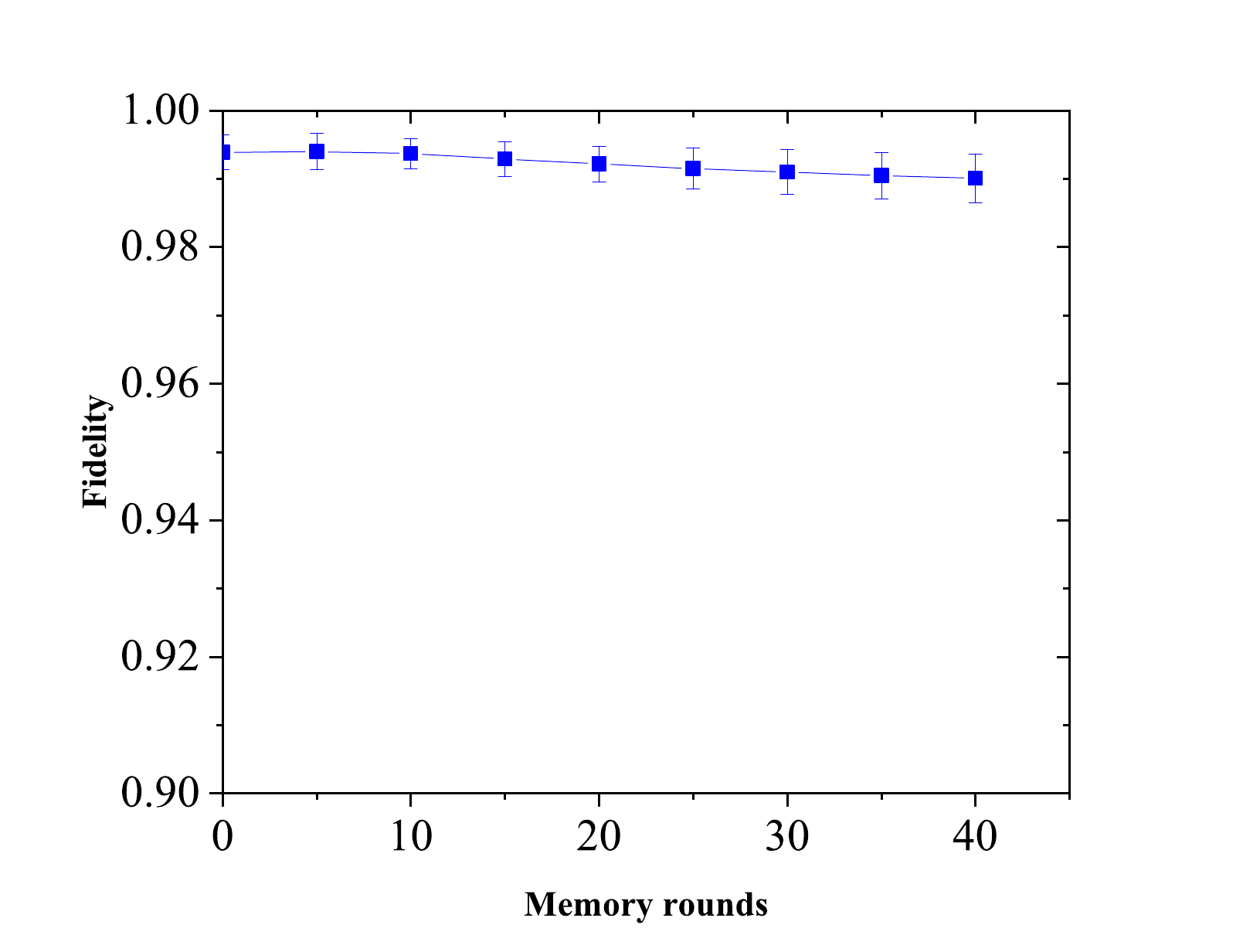}
	\caption{State fidelity of time-bin qubit states after storing N rounds in the optical cavity. The photon counts of one single measurement are collected for 1 s with 1 ns detection window.}
	\label{Fig1}
\end{figure}
Table 1 provides an overview of some typical quantum memory works, highlighting key representative indicators such as storage time, efficiency, fidelity, and bandwidth. This comparison enables us to have a more intuitive and clear understanding of the characteristics and differences of these quantum memory works.
\begin{table}[H]
	\centering
	\caption {Comparison of some typical quantum memories.
	}\label{T1}
	\begin{tabular}{ccccc}
		\hline
		& Storage time & Efficiency & Fidelity & Bandwidth   \\[6pt]
		\hline	
		\cite{trap1}  & $10$ s & - & 81\%& <10MHz \\[5pt]
		\hline	
		\cite{chip4}  & $1.8$ ns & 6.3\% & -& - \\[5pt]
		\hline	
		\cite{com1}   & $550$ ns & 11.3\% & 89.6\% & 34GHz \\[5pt]
		\hline	
		\cite{com2}  & $325$ ns & 6.9\% & 98.3\% & 100MHz \\[5pt]
		\hline
		Ours  & $520$ ns & 13.5\% & 99.1\% & >1THz \\[5pt]
		\hline	
	\end{tabular}
\end{table}
\begin{figure*}[ht]
	\centering	
	\begin{overpic}[width=0.22\hsize]{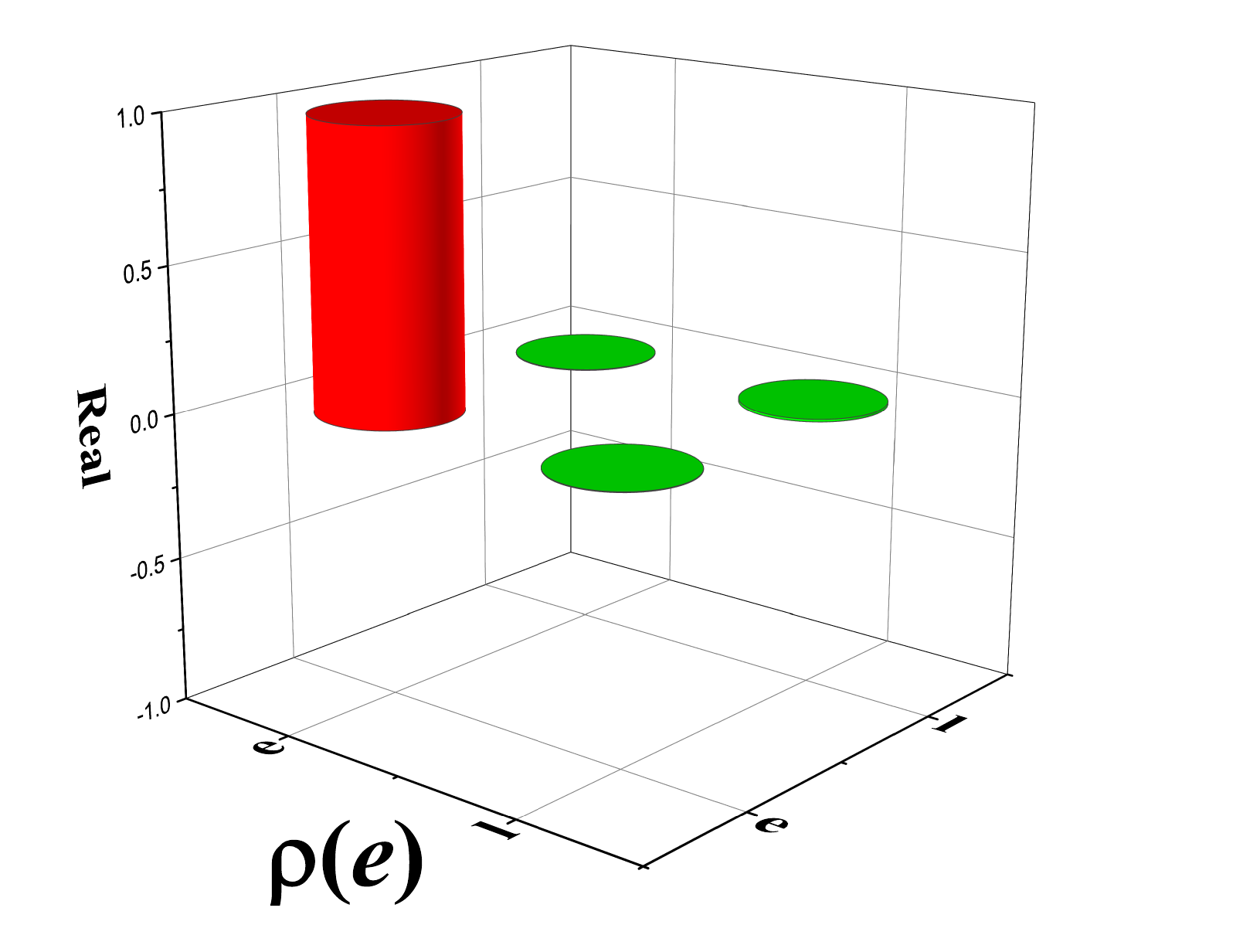}
		\put(3,70){(a)}
	\end{overpic}\hspace{0.01cm}
	\begin{overpic}[width=0.22\hsize]{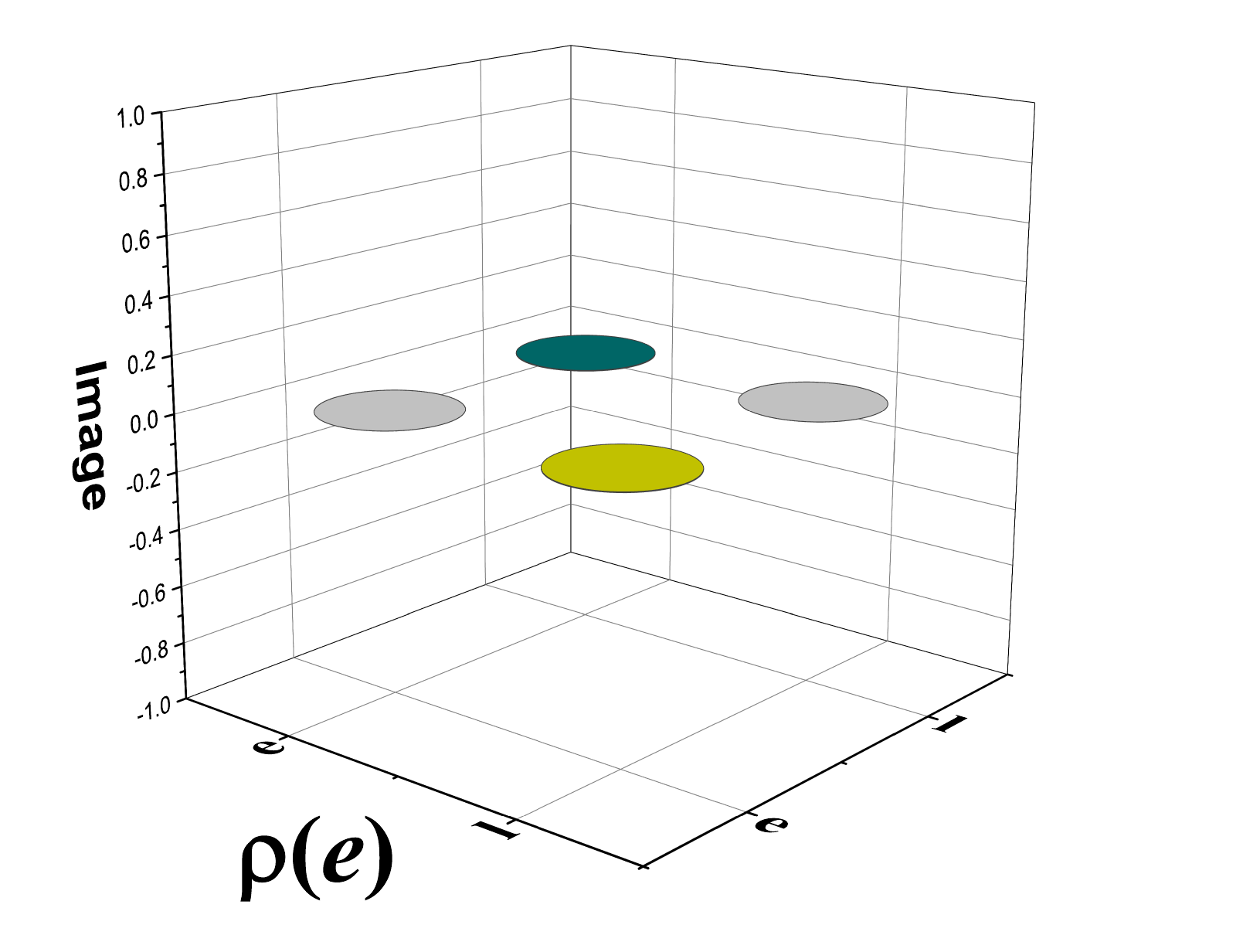}
		\put(3,70){(b)}
	\end{overpic}\hspace{0.01cm}
	\begin{overpic}[width=0.22\hsize]{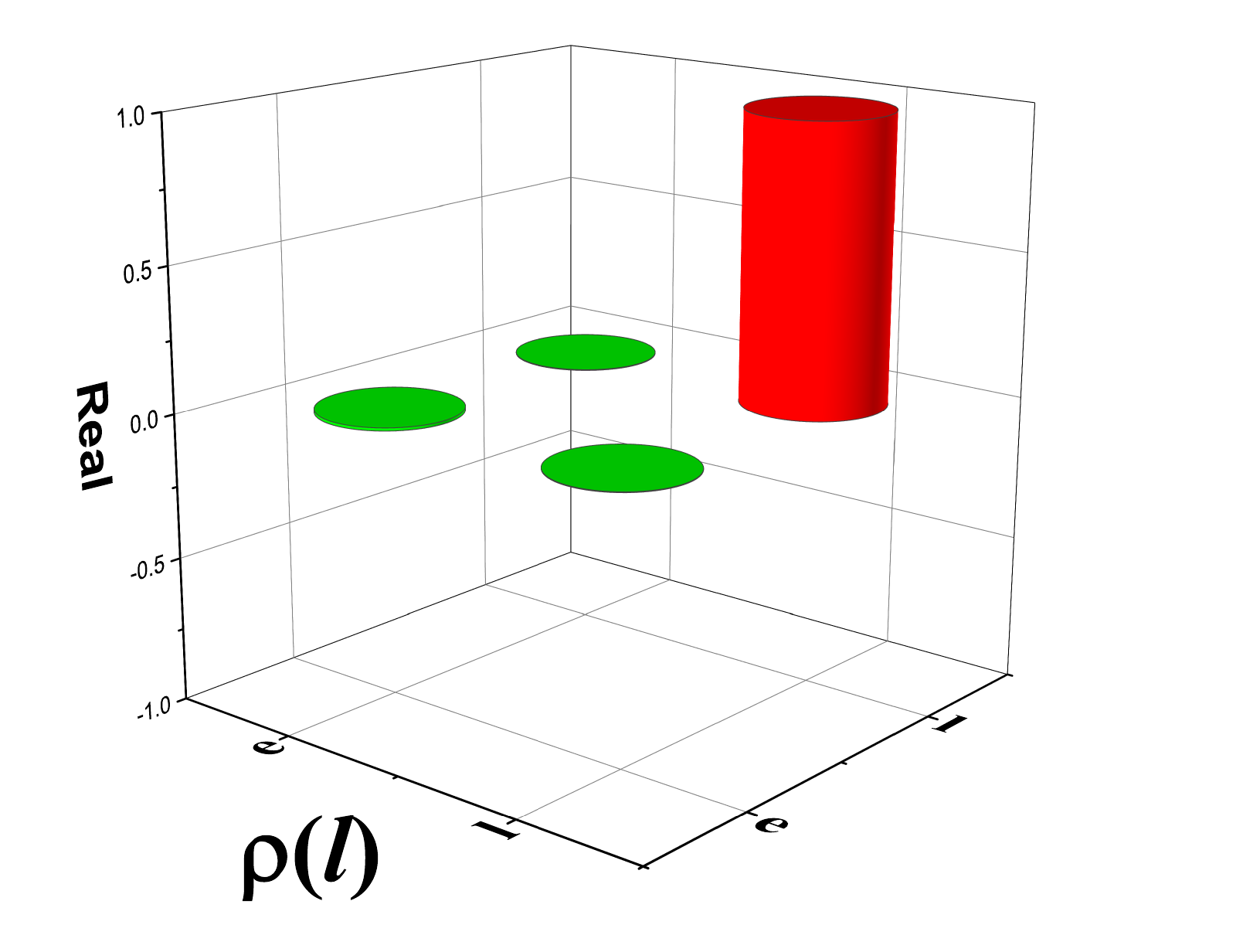}
		\put(3,70){(c)}
	\end{overpic}
	\hspace{0.01cm}
	\begin{overpic}[width=0.22\hsize]{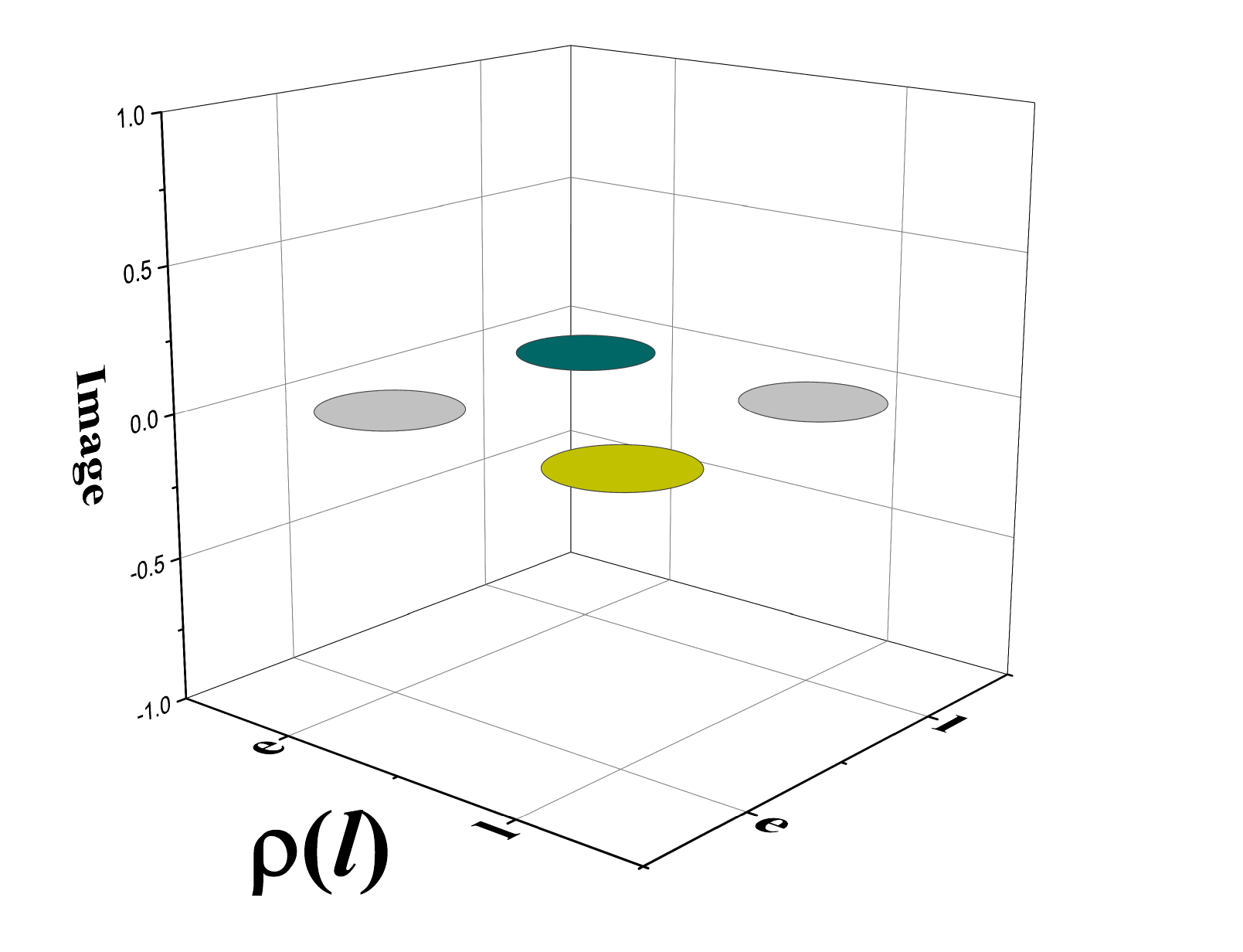}
		\put(3,70){(d)}
	\end{overpic}\hspace{0.01cm}
	\begin{overpic}[width=0.22\hsize]{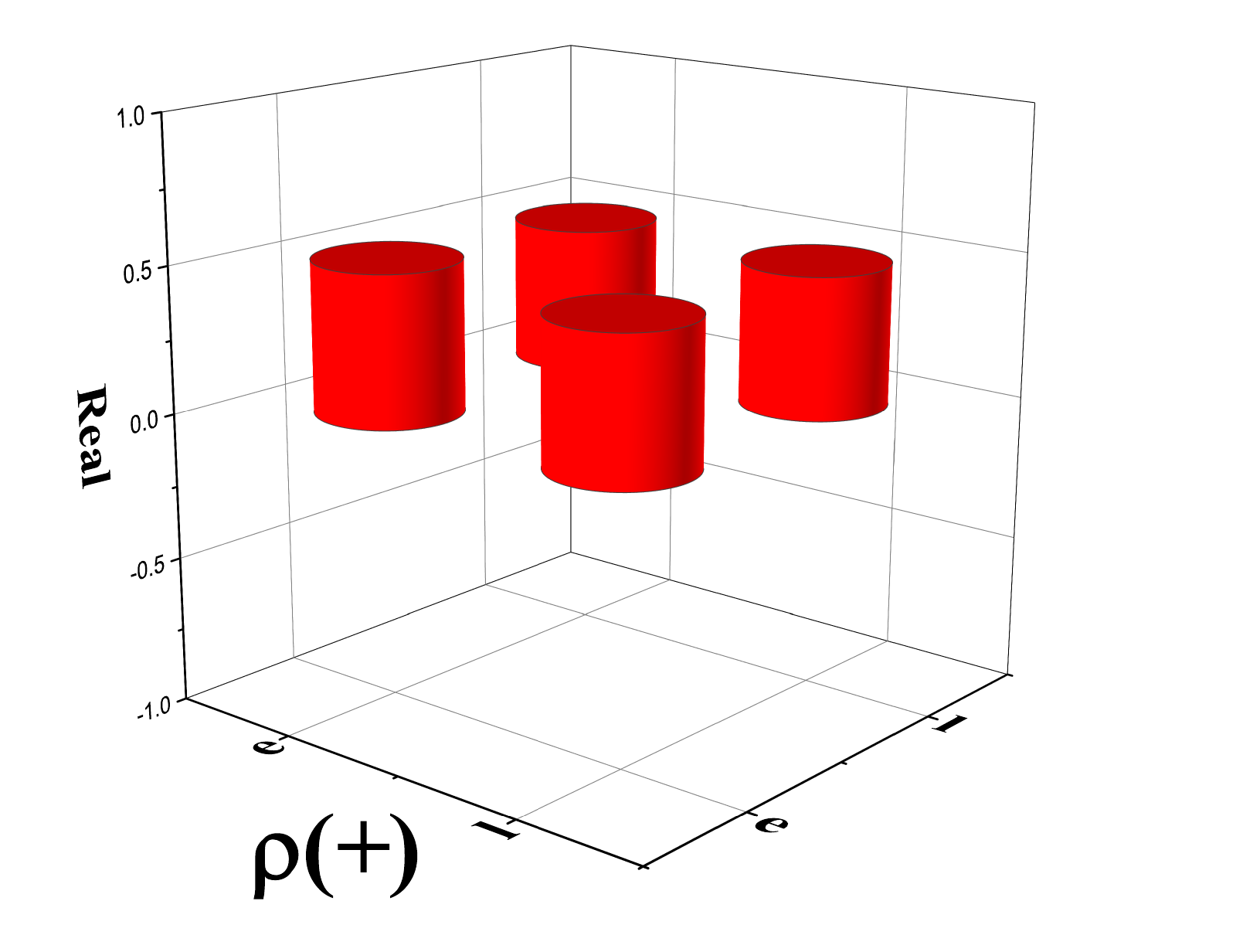}
		\put(3,70){(e)}
	\end{overpic}
	\hspace{0.01cm}
	\begin{overpic}[width=0.22\hsize]{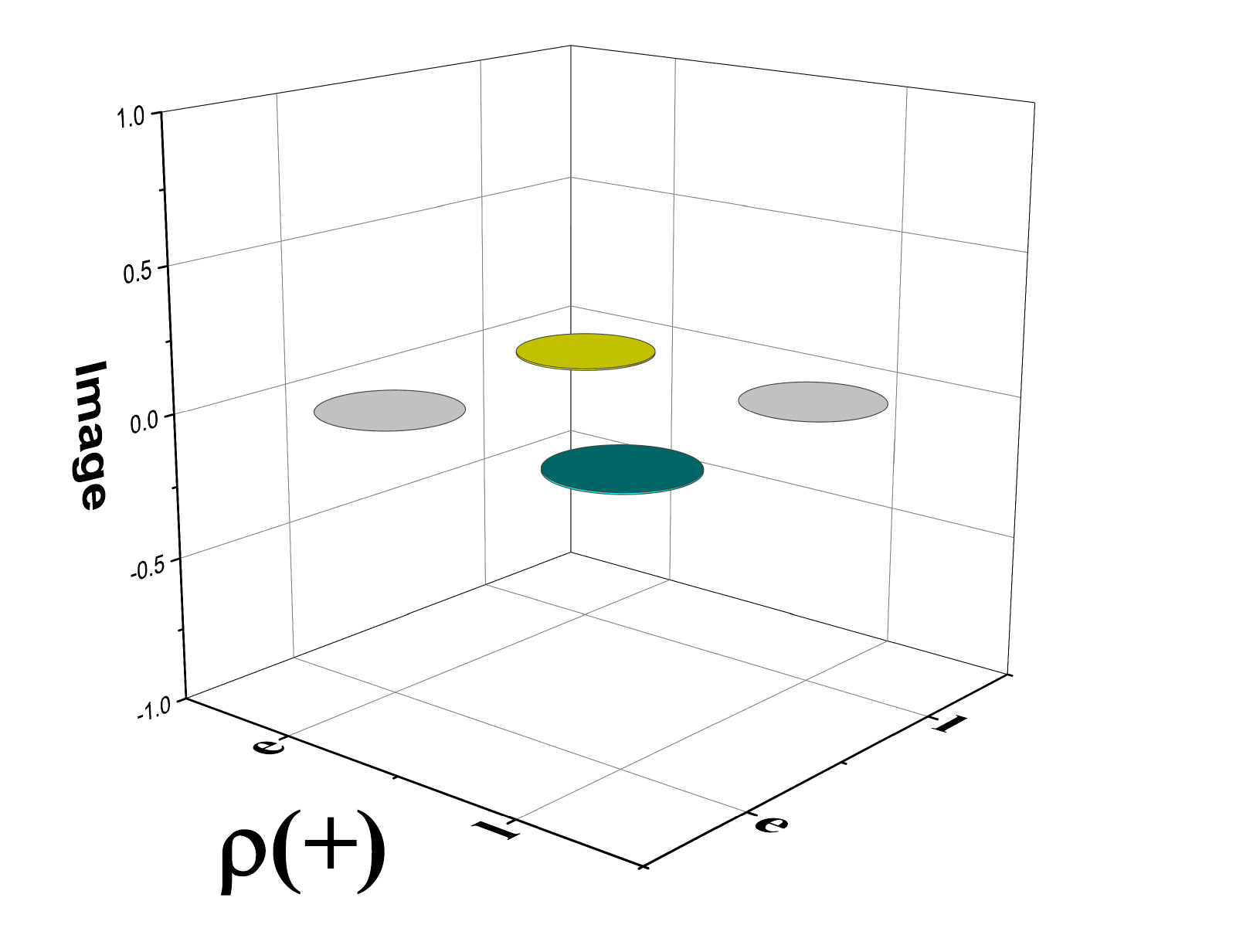}
		\put(3,70){(f)}
	\end{overpic}\hspace{0.01cm}
	\begin{overpic}[width=0.22\hsize]{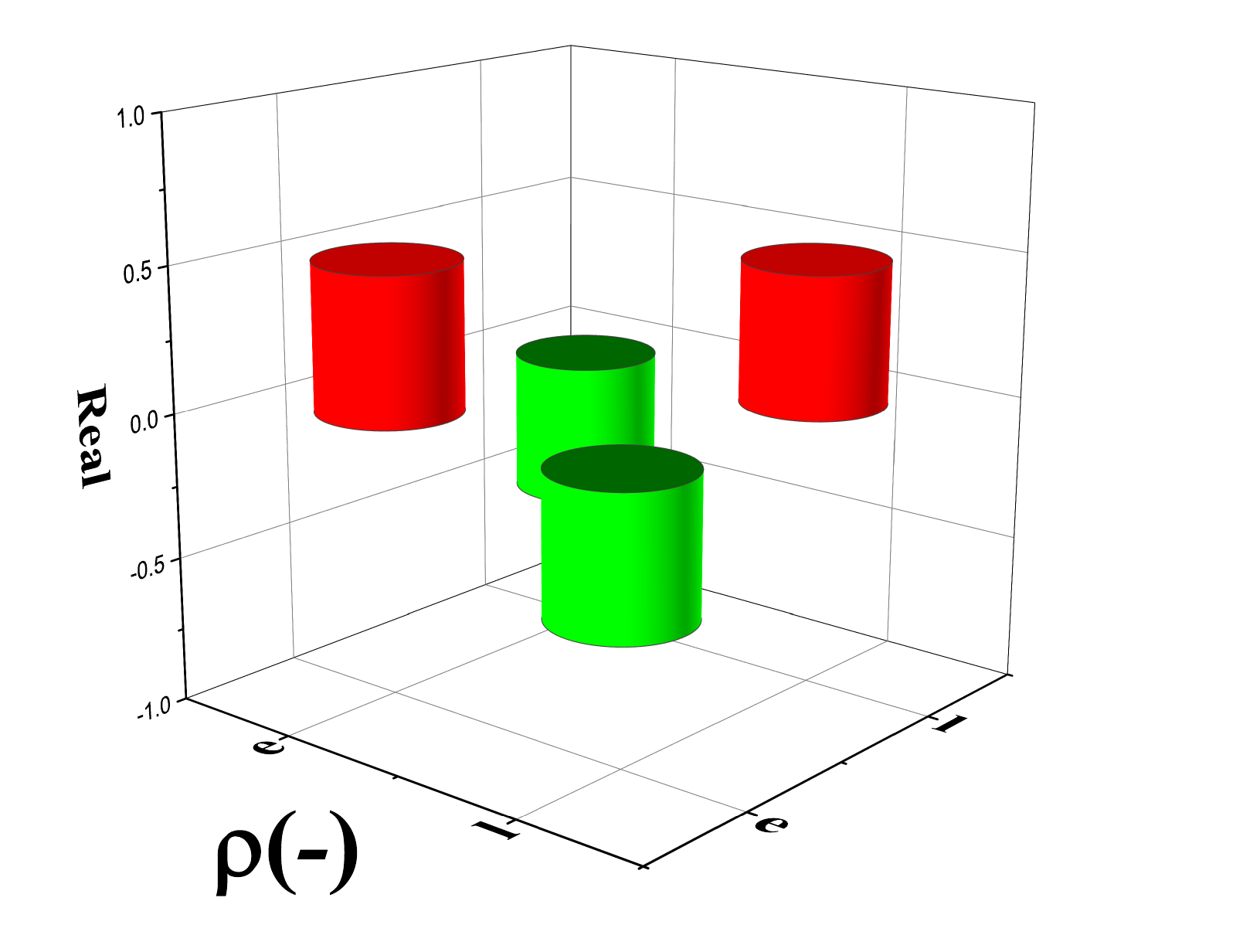}
		\put(3,70){(g)}
	\end{overpic}
	\hspace{0.01cm}
	\begin{overpic}[width=0.22\hsize]{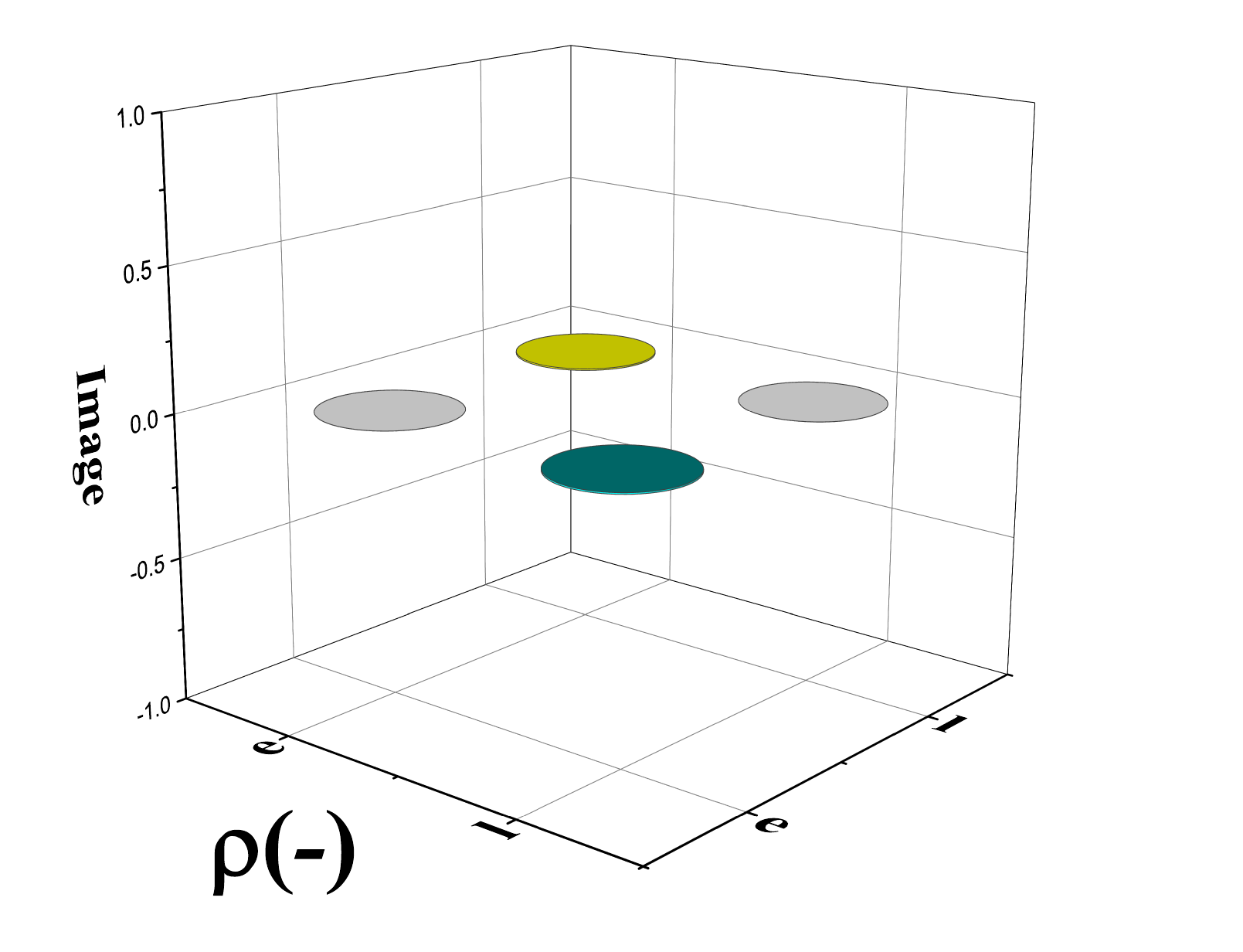}
		\put(3,70){(h)}
	\end{overpic}\hspace{0.01cm}
	\begin{overpic}[width=0.22\hsize]{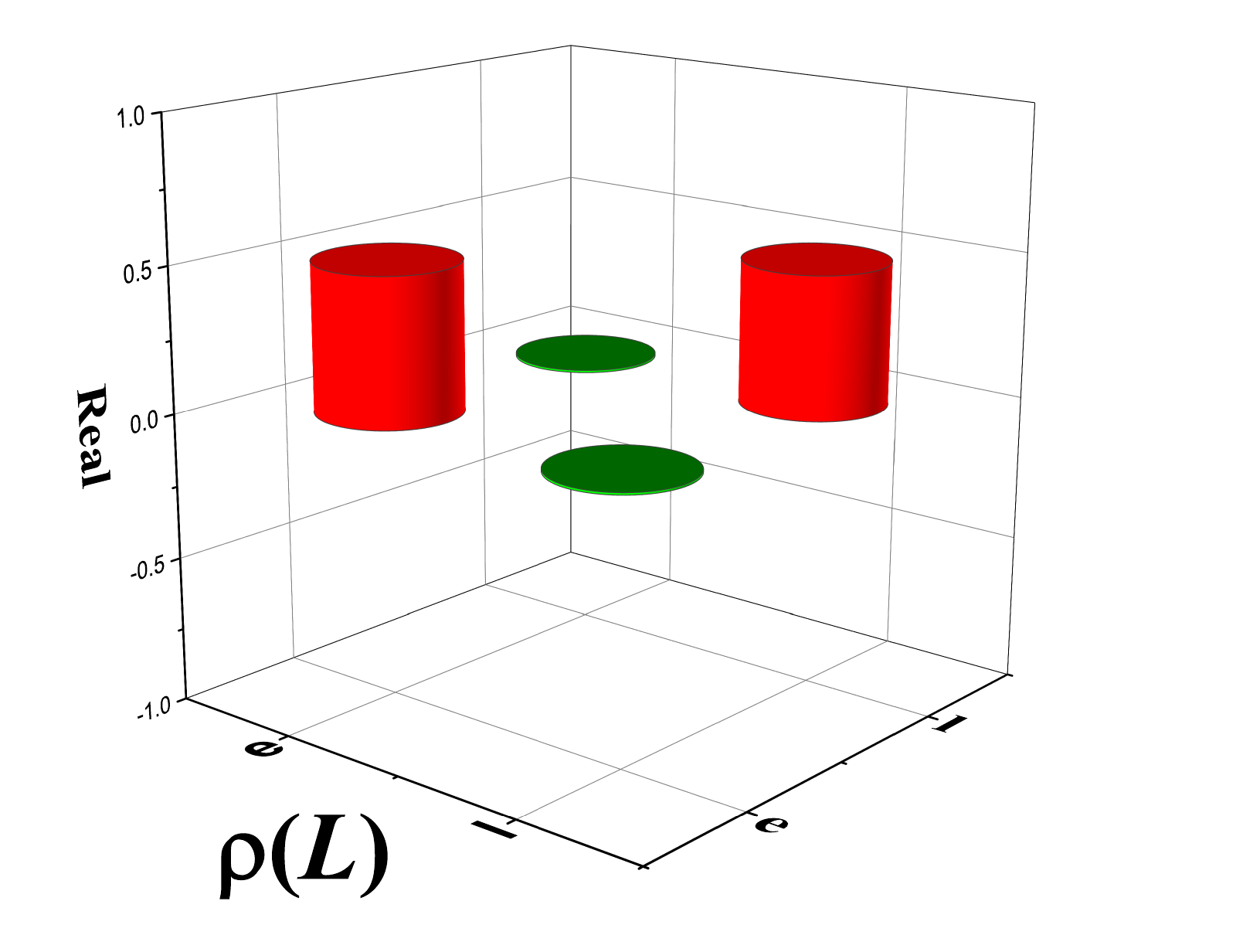}
		\put(3,70){(i)}
	\end{overpic}
	\hspace{0.01cm}
	\begin{overpic}[width=0.22\hsize]{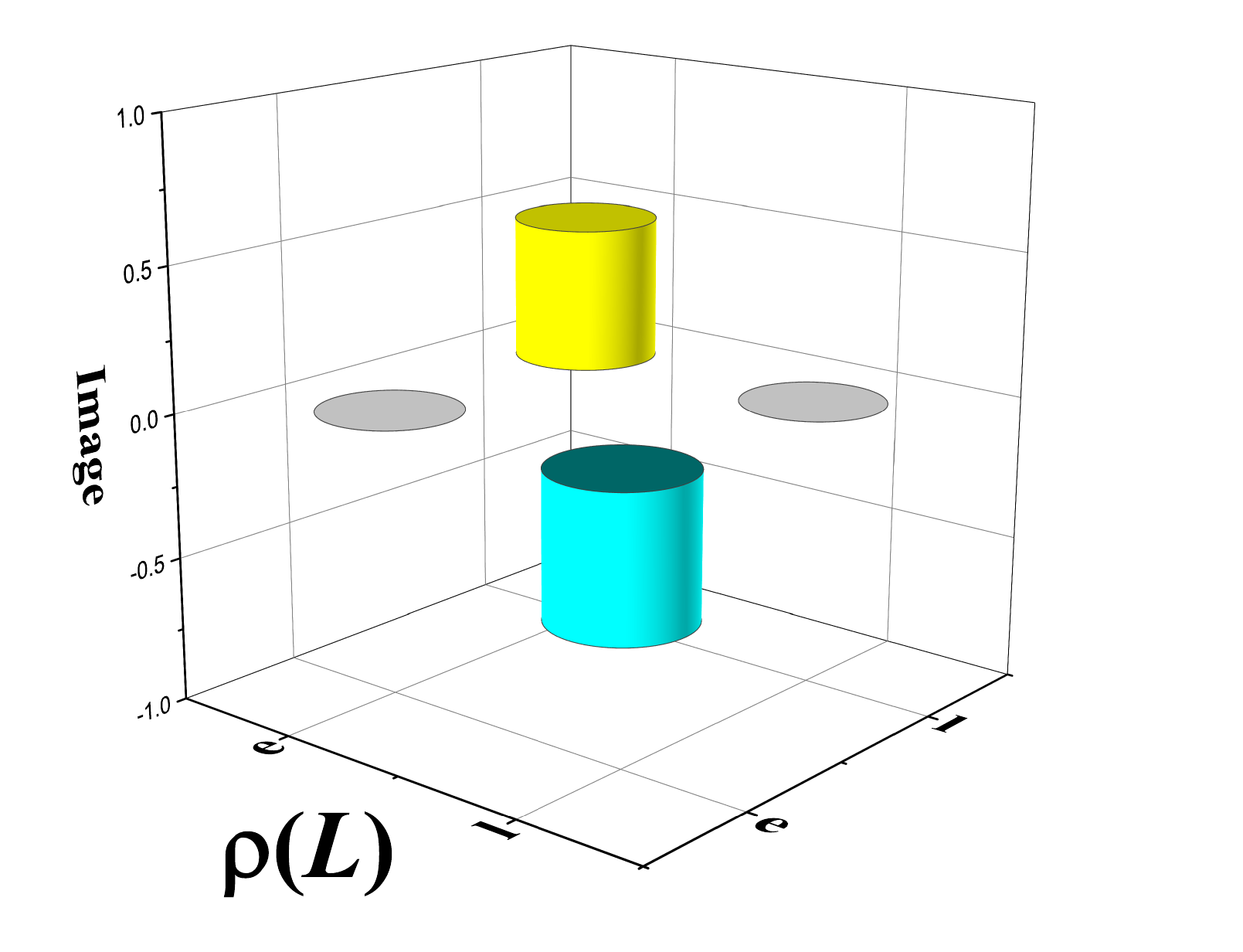}
		\put(3,70){(j)}
	\end{overpic}\hspace{0.01cm}
	\begin{overpic}[width=0.22\hsize]{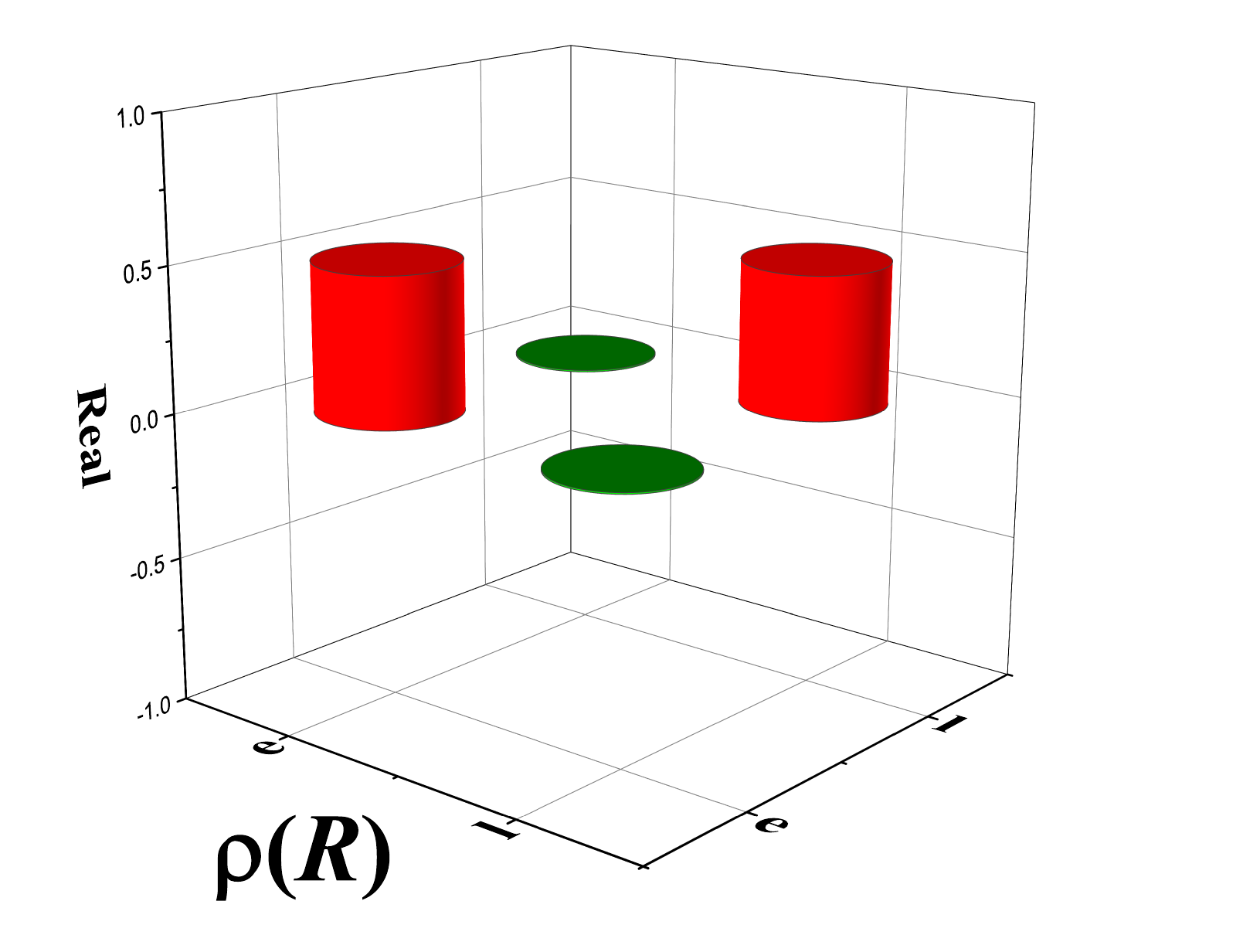}
		\put(3,70){(k)}
	\end{overpic}
	\hspace{0.01cm}
	\begin{overpic}[width=0.22\hsize]{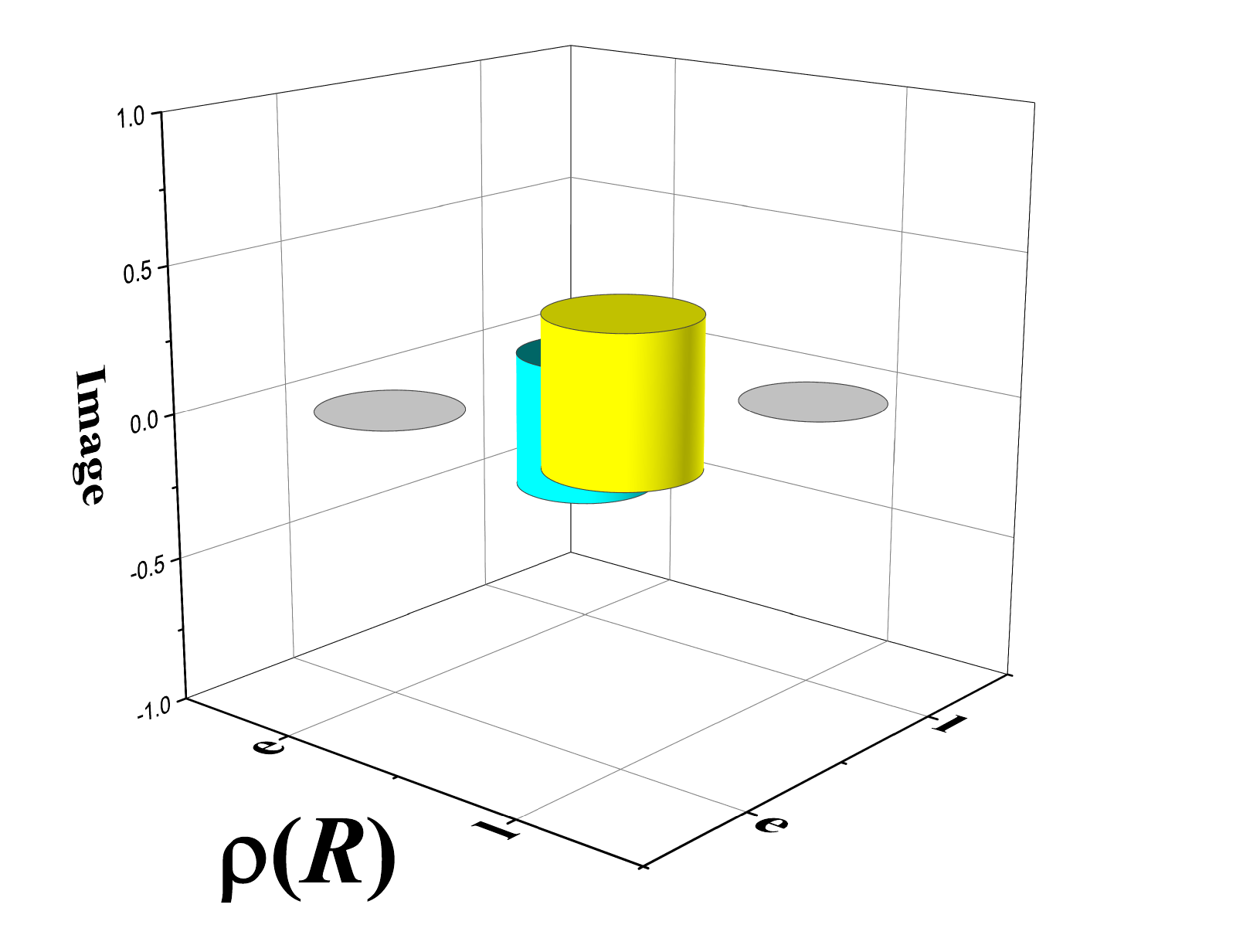}
		\put(3,70){(l)}
	\end{overpic}
	\caption{The reconstructed density matrices of the retrieved states of $|e\rangle$, $|l\rangle$, $|+\rangle$, $|-\rangle$,  $|L\rangle$ and $|R\rangle$  at the 30-th memory round. The photon counts of one single measurement are collected for 1 s with 1 ns detection window. (a),(c),(e),(g),(i),(k) Real components; (b),(d),(f),(h),(j),(l) Imaginary components.}	
	\label{Fig3}
\end{figure*}

\section{Conclusion}
In conclusion, we have experimentally realized quantum storage for time-bin qubit states with high performances. It can exhibit the single-round storage efficiency of 95.0\% and the state fidelity exceeding 99.1\%, which is very promising for the application of quantum key distribution. This accomplishment constitutes an improvement in the fields of quantum communication and quantum cryptography, offering a reliable and efficient quantum storage solution for future applications.\par

\section{ACKNOWLEDGMENTS}
The authors gratefully appreciate Dr. Zhi-Yuan Zhou for enlightening discussion during the experiment. This work is supported by Natural Science Foundation of Jiangsu Province (BE2022071), the National Natural Science Foundation of China (NSFC) (12074194, 12104240, 62101285, 62471248) and the Postgraduate Research \& Practice Innovation Program of Jiangsu Province (KYCX220954).\par
\section{COMPETING INTERESTS}
The authors declare that there are no competing interests.

\end{document}